\documentclass[aps,twocolumn,floats,superscriptaddress,prd]{revtex4}
\usepackage{graphicx, epsfig, bm,amsmath}

\usepackage{color}
\usepackage{hyperref}
\usepackage{ifthen}
\usepackage{xstring}
\begin{document}


\newcommand{\zmin}{z_{\rm min}}
\newcommand{\zmax}{z_{\rm max}}
\newcommand{\dz}{\Delta z}
\newcommand{\dzsub}{\Delta z_{\rm sub}}
\newcommand{\zminsn}{z_{\rm min}}
\newcommand{\nzpc}{N_{z,{\rm PC}}}
\newcommand{\amax}{a_{\rm max}}
\newcommand{\atr}{a_{\rm tr}}
\newcommand{\aeq}{a_{\rm eq}}

\newcommand{\wmin}{w_{\rm min}}
\newcommand{\wmax}{w_{\rm max}}
\newcommand{\wfid}{w_{\rm fid}}

\newcommand{\lcdm}{$\Lambda$CDM}
\newcommand{\hmpc}{h^{-1}\,{\rm Mpc}}

\newcommand{\om}{\Omega_{\rm m}}
\newcommand{\ode}{\Omega_{\rm DE}}
\newcommand{\ok}{\Omega_{\rm K}}
\newcommand{\omhh}{\omega_{\rm m}}
\newcommand{\olhh}{\omega_{\rm \Lambda}}

\newcommand{\winf}{w_{\infty}}
\newcommand{\scrm}{\mathcal{M}}
\newcommand{\osf}{\Omega_{\rm sf}}
\newcommand{\omf}{\Omega_{\rm m}^{\rm fid}}
\newcommand{\scrmf}{\mathcal{M}^{\rm fid}}
\newcommand{\rhode}{\rho_{\rm DE}}
\newcommand{\rhoc}{\rho_{{\rm cr},0}}

\newcommand{\dlum}{d_{\rm L}}
\newcommand{\dlss}{D_*}
\newcommand{\mpl}{M_{\rm Pl}}
\newcommand{\ds}{\displaystyle}

\newcommand{\gpr}{G^{\prime}}

\newcommand{\mmcomment}[1]{\textcolor{red}{[{\bf MM}: #1]}}
\newcommand{\dhcomment}[1]{\textcolor{magenta}{[{\bf DH}: #1]}}
\newcommand{\wh}[1]{\textcolor{blue}{[{\bf WH}: #1]}}


\pagestyle{plain}

\title{Hiding dark energy transitions at low redshift}

\author{Michael Mortonson}
\affiliation{Kavli Institute for Cosmological Physics 
       and Department of Physics,
        University of Chicago, Chicago, IL 60637}

\author{Wayne Hu}
\affiliation{Kavli Institute for Cosmological Physics 
       and Department of Astronomy \& Astrophysics,
        University of Chicago, Chicago, IL 60637}

\author{Dragan Huterer}
\affiliation{Department of Physics, University of Michigan, 
450 Church St, Ann Arbor, MI 48109-1040}

\begin{abstract}
  We show that it is both observationally allowable and theoretically possible
  to have large fluctuations in the dark energy equation of state
  as long as they occur at ultra-low redshifts $z \lesssim 0.02$.
  These fluctuations would masquerade as a local transition in the Hubble rate
  of a few percent or less and escape even future, high precision, high
  redshift measurements of the expansion history and structure.  Scalar field
  models that exhibit this behavior have a sharp feature in the potential that
  the field traverses within a fraction of an $e$-fold of the present.  The
  equation of state parameter can become arbitrarily large if a sharp dip or
  bump in the potential causes the kinetic and potential energy of the field
  to both be large and have opposite sign.  While canonical scalar field
  models can decrease the expansion rate at low redshift, increasing the local
  expansion rate requires a non-canonical kinetic term for the scalar field.
\end{abstract}
\maketitle


\section{Introduction}
\label{sec:introduction}

With the ever-tightening constraints on the acceleration of cosmic expansion
(e.g.~\cite{FriTurHut}), it is interesting to ask whether the measurements are
still compatible with any substantial deviations from a cosmological constant.
While it is well known that at high redshifts ($z \gg 1$) order unity deviations
may still exist in the dark energy equation of state, this is during an epoch where 
dark energy has very little impact on the expansion
history.

Interestingly, the other place where an order unity transition in the dark
energy equation of state can be hidden from data is in the local universe, at
ultra-low redshifts.  Indeed there have been hints from Type Ia supernova (SN)
data that there may be a discontinuous break in the Hubble diagram (a ``Hubble
bubble'') of $\sim 5$\% at a redshift of $z\sim 0.023$
\cite{Zehavi_bubble,Jha_bubble}.  In the $\Lambda$CDM paradigm, a
discontinuity could be explained by a local void, but its amplitude would have
to be atypically large \footnote{We do {\it not} address  whether a
  much larger void out to $z \sim 0.1$ can make the exterior Hubble parameter
 small enough to be compatible with  no dark energy \cite{Alexander07}.}.  Moreover, these findings have
been brought into question with the advent of new SN data and expanded studies
of SN systematics, particularly color corrections
\citep{Lifan_Wang_07,Conley_bubble,Constitution,SHOES}.

In this {\it Brief Report}, we consider the local measurements as an upper
limit on recent variations in the dark energy density.  In \S \ref{sec:hiding},
we show how such sudden transitions are hidden from current and future high
redshift measurements, determine the implied requirements on dark energy, and
construct explicit scalar field models of the transition.  We discuss these
results in \S \ref{sec:discussion}.

\section{Hiding Dark Energy Transitions}
\label{sec:hiding}

\subsection{Low Redshift Transitions}

Large transitions in the dark energy density can evade current observations
if they only affect low redshifts $z_t \ll 0.1$.  Galaxy surveys and the growth
of structure are largely insensitive to such transitions simply because the
enclosed volume and number of $e$-folds of the expansion are too small.

Distance measures are affected at all redshifts, but only have
measureable changes at very low redshifts. 
For example, cosmic
microwave background (CMB) and baryon acoustic oscillation (BAO) distances out
to higher redshifts are largely unaffected since the shift is a small
fraction of the total distance, $\delta \dlum \sim -(z_t /H_0) \delta
H_0/H_0$.

Given CMB and BAO absolute distance measures, one might expect their
relationship to SN distance measures at $z\gg z_t$ to be affected by a dark 
energy transition.  
SN data measure the relative luminosity distance
$\dlum$ between SNe in the sample, $\dlum(z)/\dlum(z_{\rm min})$, where $z_{\rm
  min}$ is the minimum SN redshift in the survey.  Ordinarily, one would
assume the Hubble law
\begin{equation}
\lim_{z_{\rm min} \rightarrow 0} \dlum(z_{\rm min})= {z_{\rm min} \over H_0}
\label{eqn:hubblelaw}
\end{equation} 
and call $H_0\dlum(z)$ the observable SN distance.  This measure would seem to
be sensitive to local variations in the expansion rate when combined with $\dlum(z)$
from the CMB and BAO.  However, if the dark
energy density
undergoes a transition at $z_t < z_{\rm min}$, the expansion rate is no longer
constant in $z$ at $z<z_{\rm min}$, leading to departures from a pure Hubble law.

 In other
words, apparent SN magnitudes
\begin{equation}
m(z)=5\log[H_0\dlum(z)]+(M-5\log H_{0}+25)\,
\label{eq:apparent_mag}
\end{equation} 
are unaffected by local dark energy transitions.
The quantity in parentheses is an unknown constant 
involving the absolute SN magnitude $M$.  Since $\dlum(z\gg z_t)$ is essentially unchanged
 if $H_0$ jumps in value 
locally, the observable $m(z)$ remains unchanged.   In fact, the most precise
measurement of $H_0$ to date uses a maser--Cepheid calibration of absolute 
SN distances and $M$ {\it above} $z_{\rm min}=0.023$ \cite{SHOES}. 
Only distance measurements below $z_t \lesssim 0.02$ would be sensitive to such a jump
and current data limit its amplitude to be $\lesssim 5\%$.

\subsection{Dark Energy Requirements}

Let us determine what is required of dark energy to achieve such a Hubble
transition.  Consider a scenario in which the true Hubble constant is given by
\begin{equation}
H_0 \approx (1+\delta) \tilde H_0\,,
\end{equation} while 
 the high-redshift expansion rate is left unchanged. 
Here and throughout tildes denote values in a
flat $\Lambda$CDM reference model.

 To achieve this let
us take a dark energy density of the form
\begin{equation}
\rho_{\rm DE}(z) =[1+ f(z)] \tilde \rho_{\Lambda}
\label{eqn:hubblemodel}
\end{equation}
and demand that
\begin{equation}
f(z) = \begin{cases}
0& z\gg z_{\rm t}\,, \\
{2\delta / \tilde\Omega_{\Lambda}} & z = 0 \,,
\end{cases}
\label{eq:flimits}
\end{equation}
where $z_t$ is the transition redshift. Since we have not altered
the physical matter density, $\Omega_{\rm m} h^2 = \tilde \Omega_{\rm m} \tilde
h^2 $ and
\begin{equation}
\Omega_{\rm m} = {\tilde\Omega_{\rm m}
\over 1+f(0) } ={1-\tilde\Omega_{\Lambda} \over 1+2\delta/\tilde \Omega_\Lambda}= 1-\Omega_{\rm DE}\,.
\end{equation}
Furthermore, CMB constraints on the matter density at
recombination are automatically satisfied.

Given these requirements, three general features of a dark energy Hubble
transition remain to be specified.  The first two, the transition redshift
$z_t$ and the duration of the transition $\Delta z$,
stringently constrain  the equation of state
\begin{equation}
1+w = {1\over 3} {(1+z) f' \over 1+f}\,,
\label{eq:one_plus_w}
\end{equation}
where $f'\equiv df/dz$, given our requirements  that
$z_t \ll 1$ and $\Delta z < z_t$.  The average equation of state at low
redshifts is $1+w \sim -\delta / \Delta z$.  Thus a transition at $z_t
\lesssim 0.02$ with an amplitude greater than a few percent requires an exotic
equation of state which deviates from a cosmological constant by more than
order unity.  Moreover, transitions with $0<\delta\ll 1$ require phantom
equations of state with $w < -1$.

The remaining freedom is somewhat more subtle.  Although the dark energy
density, Hubble parameter, and distance--redshift relation are specified by
our description so far, the dark energy equation of state {\it after} the
transition (i.e.\ at $z \ll z_t$) is not.  Given the small
fraction of an $e$-fold of expansion between the transition and the present,
any post-transition $1+w$ that is order unity or less would give the same
cosmological observables.  This freedom in $w(z\ll z_t)$ allows one to build many
models that produce a given Hubble transition.

\begin{figure}[t]
\centerline{\psfig{file=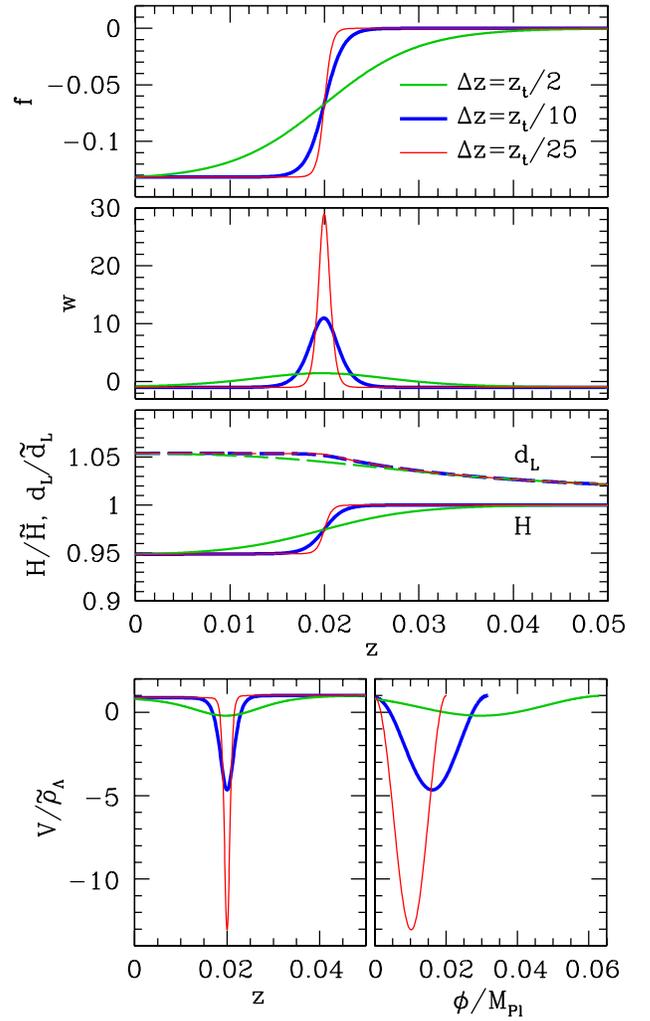, width=3.2in}}
\caption{ Hubble transitions with different redshift widths corresponding to
  the potential-dominated model of Eqs.~(\ref{eq:V_rec})
  and~(\ref{eqn:simplemodel}): $\Delta z = z_t/2$ (\emph{medium thickness,
    green}), $z_t/10$ (\emph{thick, blue}), and $z_t/25$ (\emph{thin,
    red}). All models have a transition at $z_t=0.02$ with amplitude
  $\delta=-0.05$ and tildes denote the $\Lambda$CDM reference model with
  $\tilde \Omega_{\rm m}=0.24$, $\tilde \Omega_{\Lambda}=0.76$, and $\tilde
  h=0.73$.  \emph{Upper panels}: redshift evolution of various observables.
  \emph{Lower panels}: scalar field model potential.
}
\label{fig:models2}
\end{figure}

\subsection{Dark Energy Models}

Let us construct scalar field models that satisfy the Hubble transition
requirements.  For a canonical kinetic term, we can take the dark energy
density and equation of state from Eqs.~(\ref{eqn:hubblemodel}) 
and~(\ref{eq:one_plus_w}) and
reconstruct the scalar field potential
\cite{Starobinsky,reconstr,Saini_reconstr,Nakamura_Chiba}.  Since the
potential energy is $V = (\rho_{\rm DE} - p_{\rm DE})/2$, we have
\begin{eqnarray}
{V(z)} &=& {1\over 2}{(1-w(z))}\rho_{\rm DE}(z) \nonumber\\
&=&[(1+f)-(1+z)f'/6] \tilde \rho_{\Lambda}  \,.
\label{eq:V_rec}
\end{eqnarray}
The kinetic
energy of the field is $\dot \phi^2/2 = (\rho_{\rm DE} + p_{\rm DE})/2$, so
\begin{eqnarray}
\phi(z) &=&  
\int_0^z |(1+w(z'))\rho_{\rm DE}(z')  |^{1/2}
{dz'\over (1+z')H(z')} \nonumber\\
&=&\sqrt{ \tilde\rho_\Lambda\over 3} \  \int_0^z  \left( {| f'  |\over 1+z'} \right)^{1/2}  {dz'\over H(z')}\,,
\label{eq:phi_rec} 
\end{eqnarray}
where, without loss of generality, we have taken the sign of the field to be
positive and set its present value to zero.  From the two equations above
one can implicitly get $V(\phi)$.  During the transition, the field in 
units of the reduced Planck mass $\mpl \equiv (8\pi G)^{-1/2}$ rolls a distance 
$\Delta \phi/\mpl \sim |\delta \Delta z|^{1/2}$ which for typical values
gives $10^{-2}$.

Implicit in this construction is the requirement that the kinetic energy
$\rho_{\rm DE} + p_{\rm DE}$ remain a positive quantity, which implies that
$f$ must monotonically increase with $z$.  Note that if $\rho_{\rm DE}$ can
switch signs, this differs from the requirement that $1+w>0$.  A change of
sign in both the dark energy density and $1+w$ can occur if $f<-1$ but
requires such a large change in $H(z)$ that $\delta H_0^2/\tilde H_0^2
\lesssim -\tilde \Omega_{\Lambda}$, i.e.\ almost to the point that the
expansion becomes a contraction at low redshift.  For all models with
canonical kinetic terms, the requirement that $f'>0$ combined with the
restrictions of Eq.~(\ref{eq:flimits}) implies that $\delta<0$.

To construct $\delta >0$ models we require a non-canonical kinetic term.  The
simplest possibility is to just reverse the sign of the kinetic term.  The
expressions in Eq.~(\ref{eq:V_rec}) and~(\ref{eq:phi_rec}) are then identical, 
but the scalar field now rolls up the potential.

\begin{figure}[t]
\centerline{\psfig{file=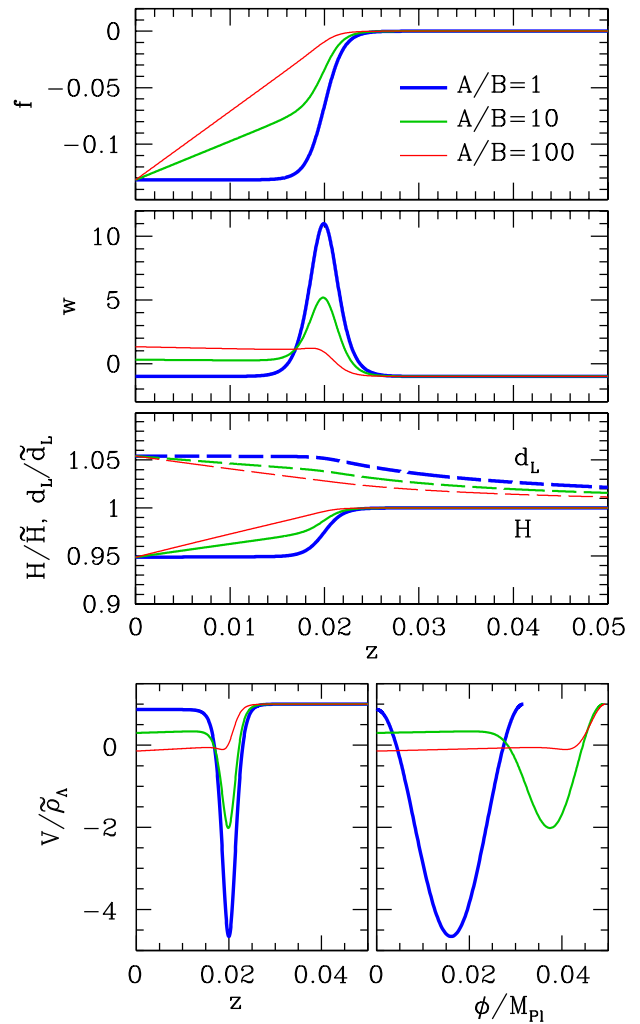, width=3.2in}}
\caption{
  Hubble transitions with different ratios of potential to kinetic
  energy at $z=0$, using the generalized model of Eqs.~(\ref{eq:kinv})
  and~(\ref{eq:kinf}) with $A/B = 1$ (\emph{thick, blue}), $10$ (\emph{medium thickness, green}), and
  $100$ (\emph{thin, red}). The $A/B=1$ model corresponds to zero  kinetic energy
  after the transition as in Fig.~\ref{fig:models2}, and $A/B \rightarrow \infty$ to maximal
  kinetic energy.
 All models have $z_t=0.02$ of width $\Delta z =
  z_t/10$ and amplitude $\delta=-0.05$.  }
\vskip -0.3cm
\label{fig:models1}
\end{figure}

Even given these requirements, there are many scalar field
potentials that can reproduce a given $\delta$.  Let us start with the
assumption that the scalar field on the low redshift side of the transition
becomes potential energy dominated directly after the transition.  This
implies that $f(z)$ strictly approaches a constant for $z\ll z_t$.  For
example, we can take
\begin{eqnarray}
f(z) &=& {2\delta \over \tilde\Omega_{\Lambda}}{S(z) \over S(0)} \,,\nonumber\\
S(z) & = & {1\over 2} \left[1 - \tanh\left({ z-z_{\rm t} \over \Delta z}\right)\right] \,.
\label{eqn:simplemodel}
\end{eqnarray}
The scalar field potentials reconstructed from this assumption are shown in
Fig.~\ref{fig:models2}.  Note that to achieve the $|1+w|>2$ equation of state
for a large amplitude, rapid transition, we require {\it negative} potentials
where a large positive kinetic energy and large negative potential energy
cancel to leave a small total energy (see Fig.~\ref{fig:models2}).  In fact,
the potential can become arbitrarily negative and $|1+w|$ arbitrarily large
without measurably changing the main observable $d_L(z)$.  This is because no
matter how sharp the transition in the expansion rate $H(z)$, distances are
always a smooth function of redshift.

Models that leave the field with more kinetic energy after the transition
are also possible.  For example, consider the class of potentials defined by
\begin{equation}
V(z)/\tilde\rho_{\Lambda}-1 = A {S(z) \over S(0)} - B{(1+z)\over 6} {S'(z)\over S(0)}  \,.
\label{eq:kinv}
\end{equation}
Our potential-dominated model in Eq.~(\ref{eqn:simplemodel}) corresponds to
$A=2\delta/\tilde\Omega_\Lambda$, $B = A$.  After the transition, the $A$ term
dominates and therefore sets the level of $V(z=0)$ (see
Fig.~\ref{fig:models1}).  The corresponding model for $f(z)$,
obtained by inverting Eq.~(\ref{eq:V_rec}), is
\begin{eqnarray}
f(z) &=& 6  \int_z^\infty dz'  {(1+z)^6 \over (1+z')^{7}} [V(z')/\tilde\rho_{\Lambda}-1] \label{eq:kinf}\\
&\approx& B {S(z) \over S(0)} + 3(A-B) \ln [1 + e^{2(z_t-z)/\Delta z}] \Delta z \,,
\nonumber
\end{eqnarray}
which implicitly defines $\delta$ through $f(0)$.  The approximation assumes
that $z_t \ll 1$.

Note that a sharp change in $V(z)$ is not sufficient to induce
a sharp change in $f(z)$.  For example, a step function potential is included
in the class of Eq.~(\ref{eq:kinv}) with ($B=0$, $\Delta z\rightarrow 0$) and has
a smooth transition in $f$ that is linear in $z$ out to $z_t$ (see
Fig.~\ref{fig:models1}).

Finally, the $\delta>0$ phantom models have identical behavior except for a
change in the sign of $(V-\tilde\rho_{\Lambda})$ and so we do not illustrate
them separately.

\section{Discussion}
\label{sec:discussion}

We have shown that it is both observationally allowable and theoretically
possible to have arbitrarily large fluctuations in the equation of state of
dark energy as long as they occur at ultra-low redshifts.  The possibility
of such fluctuations that are hidden from data creates degeneracies that are
important to understand in model-independent analyses of the dark energy
constraints \cite{PaperI}.

These fluctuations in $w(z)$ would appear as a local transition in the Hubble rate.  So
long as this change is of order a few percent or less at $z \lesssim 0.02$ it
would escape current observational constraints.  Moreover, as long as the
transition is from a constant high-redshift dark energy density, it would be
practically indistinguishable from a cosmological constant for even future
high precision distance and growth of structure measurements at high redshift.
On the other hand, future percent-level Hubble constant measurements could
place stronger limits on such transitions but will require accurate modeling
of peculiar velocities (e.g.~\cite{Hui_Greene}).

Although theoretically possible with scalar field dark energy, a Hubble
transition of this sort requires some unusual properties.  First, to
make even a percent-level change in the expansion rate over the low redshifts
in question the average equation of state {\it must} deviate by order unity
from a cosmological constant.  Moreover, models with very rapid transitions
require $|1+w| \gg 1$.  This can be achieved in scalar field models where the
potential and kinetic energy are of opposite sign and nearly cancel.  Scalar
field potentials that realize these properties have a sharp feature that must
coincidentally be traversed within a fraction of an $e$-fold of the present
epoch.

In fact, $|1+w|$ can be made arbitrarily large during the transition without a
readily observable 
effect since the transition in the distance--redshift relation remains
smooth.  Nonetheless, the absence of order unity changes in the expansion rate
in the data rules out a transition that is large enough to switch the sign of the
dark energy density and make $w(z)$ diverge before crossing $w
= -1$.  Finally, to obtain {\it enhancements}
of the low-redshift expansion rate (such as those suggested by some recent
supernova data), the scalar field must in addition have a non-canonical
kinetic term so that $1+w<0$.

The utility of studying the low redshift end of the Hubble diagram for dark
energy extends beyond the extreme context taken here of hiding order unity
transitions from current observations.  More generally, while intermediate
redshift measurements from BAO can largely take the place of local $H_{0}$
measurements for dark energy models that evolve smoothly near the present
\cite{Eisenstein:1998tu,Hu_standards}, precision measurements of the low
redshift end of the Hubble diagram not only help to constrain such smooth
models (e.g.\ \cite{Linder_lowz}), but also offer the only empirical way to
test whether dark energy has undergone recent variations in its equation of
state.

\smallskip {\it Acknowledgments:} We thank the Aspen Center for Physics where
part of this work was completed.  MM and WH were supported by the KICP under
NSF contract PHY-0114422.  MM was additionally supported by the NSF GRFP; WH by DOE contract DE-FG02-90ER-40560 and the Packard Foundation; DH by the DOE OJI grant under contract
DE-FG02-95ER40899, NSF under contract AST-0807564, and NASA under contract
NNX09AC89G.

\vfill
\bibliographystyle{arxiv_physrev}
\bibliography{hubtrans}

\end{document}